\documentclass[a4paper,fleqn,12pt]{article}
\usepackage{amsmath}
\usepackage{amssymb}
\usepackage{graphicx}
\usepackage{color}
\usepackage{cite}
 \usepackage{booktabs}

\begin{document}

\title{Extended models of nonlinear waves in liquid with gas bubbles}

\author{Nikolay A. Kudryashov, Dmitry I. Sinelshchikov}
\date{National Research Nuclear University MEPhI (Moscow Engineering Physics Institute), 31 Kashirskoe Shosse, 115409 Moscow, Russian Federation}

\maketitle

\begin{abstract}
In this work we generalize the models for nonlinear waves in a gas--liquid mixture taking into account an interphase heat transfer, a surface tension and a weak liquid compressibility simultaneously at the derivation of the equations for nonlinear waves. We also take into consideration high order terms with respect to the small parameter. Two new nonlinear differential equations are derived for long weakly nonlinear waves in a liquid with gas bubbles by the reductive perturbation method considering both high order terms with respect to the small parameter and the above mentioned physical properties. One of these equations is the perturbation of the Burgers equation and corresponds to main influence of dissipation on nonlinear waves propagation. The other equation is the perturbation of the Burgers--Korteweg--de Vries equation and corresponds to main influence of dispersion on nonlinear waves propagation.
\end{abstract}

\textit{Keywords:} Nonlinear waves; liquid with gas bubbles; reductive perturbation method; perturbed Burgers equation; nonlinear evolution equations.

\section{Introduction}
A liquid with gas bubbles has many applications in nature, industry and medicine \cite{Nakoryakov1993,Nigmatulin1990}. Nonlinear wave processes in a gas--liquid mixture were studied for the first time in works \cite{Wijngaarden1968,Wijngaarden1972,Nakoryakov1972}. The Burgers, the Korteweg--de Vries and the Burgers--Korteweg--de--Vries equations were obtained in \cite{Wijngaarden1968,Wijngaarden1972,Nakoryakov1972} for the description of long weakly nonlinear waves. The  fourth order nonlinear evolution equation for nonlinear waves in a gas--liquid mixture were obtained in \cite{Kudryashov2010,Kudryashov2010a} taking into account an interphase heat transfer. Nonlinear waves in a liquid with gas bubbles in the three--dimensional case were considered in \cite{Kudryashov2012}. Linear waves in a gas--liquid mixture under the van Wijngaarden's theory were studied in \cite{Jordan2004,Jordan2006}. In \cite{Caflisch1985} propagation of linear waves in a liquid containing gas bubbles at finite
volume fraction was considered.

In previous studies of nonlinear waves in a liquid containing gas bubbles only the first order terms with respect to the small parameter were taken into account.  On the other hand we know that using high order terms with respect to the small parameter at the derivation of nonlinear evolution equations allows us to obtain a more exact description of nonlinear waves \cite{Kodama1978,Kodama1985,Aspe1990,Fokas1996,Zhi1997,Dullin2003,Dullin2004,Depassier2006,Randruut2008,Porubov2009,Randruut2010,Depassier2012}.
Also taking into account high order correction in equations for nonlinear waves one can reveal important physical phenomena, such as interaction between dissipative and dispersive processes in a gas--liquid mixture and its influence on waves propagation, new mechanisms of waves dispersion and dissipation. Thus, it is important to study nonlinear waves in a liquid with gas bubbles taking into account second order terms in the asymptotic expansion.

We investigate nonlinear waves in a liquid with gas bubbles taking into consideration not only high order terms with respect to the small parameter but the surface tension, liquid viscosity, intephase heat transfer and weak liquid compressibility as well. To the best of our knowledge the influence of these physical properties on nonlinear waves propagation simultaneously was not considered previously.

The aim of our work is to study long weakly nonlinear waves in a liquid with gas bubbles taking into account both high order terms in the asymptotic expansion and the above mentioned physical properties in the model for nonlinear waves. We use the reductive perturbation method for the derivation of differential equations for nonlinear waves.

We apply the concept of the asymptotic equivalence, asymptotic integrability and near--identity transformations \cite{Kodama1978,Kodama1985,Fokas1996,Dullin2003,Dullin2004} for studying nonlinear equations for long waves in a gas--liquid mixture.  Asymptotically equivalent equations obtained in this work are connected to each other by a continuous group of non--local transformations \cite{Dullin2003,Dullin2004}. These transformations are near--identity transformations. Thus, we introduce families of asymptotically equivalent equations for long weakly nonlinear waves in a liquid containing gas bubbles at quadratic order. As far as all these equations are equivalent we can use a more convenient and simple equation within this family. This equation is a normal form equation. Such approach for the investigation of nonlinear evolution equation was proposed in works \cite{Kodama1978,Kodama1985}. Near--identity transformations are often named Kodama's transformations.

We derive two new nonlinear differential equations for long weakly nonlinear waves in a liquid with gas bubbles by the reductive perturbation method. In the case of dissipation main influence  nonlinear waves are governed by the perturbation of the Burgers equation. The perturbation of the Burgers--Korteweg--de Vries equation corresponds to the main influence of dispersion on nonlinear waves propagation.

We analyze dispersion relations for both equations. Near--identity transformations are used to obtain normal forms for the above mentioned equations. We show that a normal form for the equation in the dissipative case can be linearized under a certain condition on physical parameters. It is worth noting that this condition is realizable for physically meaningful values of parameters. Analytical solution of the general dissipative equation in the form of a weak shock wave is obtained and analyzed.

Two cases of a normal form equation are analyzed provided that dispersion has the main influence. The first one is the case of negligible dissipation (purely dispersive case) where nonlinear waves are governed by the generalized Korteweg--de Vries equation \cite{Olver1984}. We show that the generalized Korteweg--de Vries equation for nonlinear waves in a liquid with gas bubbles is asymptotically equivalent to one of the integrable fifth order evolution equations that are the Lax, the Sawada--Kotera and the Kaup--Kupershmidt equations. The general form of the dispersive nonlinear evolution equation seems to be nonintegrable. However, this equation admits analytical solitary wave solutions.

The rest of this work is organized as follows. In Section 2 we give the basic system of equations for nonlinear waves in a liquid with gas bubbles. We discuss the dispersion relation for linear waves as well. The main nonlinear differential equation for long weakly nonlinear waves is obtained by the reductive perturbation method. The nonlinear waves with the main influence of dissipation are studied in Section 3. Section 4 is devoted to the investigation of nonlinear waves in the case of dispersion main influence. In Section 5 we briefly discuss our results.

\section{Main differential equation for long weakly nonlinear waves in a liquid with gas bubbles}
For studying nonlinear waves in a liquid with gas bubbles we use the homogeneous model \cite{Nakoryakov1993,Nigmatulin1990}. We consider a bubble--liquid mixture as a homogeneous media with an average pressure, an average density and an average velocity. We do not take into account interaction, formation, destruction and coalescence of bubbles. Thus, the amount of gas bubbles in the mass unit is the constant $N$. We assume that all gas bubbles are spherical. The nearly isothermal approximation \cite{Prosperetti1991} is used for the modeling of heat transfer between a gas in bubbles and a liquid. In this approximation it is supposed that the temperature of the liquid is not changed and is equal to the temperature of the mixture in the unperturbed state ($T_{0}$)\cite{Prosperetti1991}. We consider influence of the liquid viscosity only at the interphase boundary. Also we take into consideration the weak compressibility of the liquid using the Keller--Miksis equation for the description of bubbles dynamics \cite{Keller1980,Prosperetti1986}. Also we consider the one--dimensional case. In these assumptions we can use the following system of equations for the description of nonlinear waves in the liquid with gas bubbles \cite{Nakoryakov1993,Nigmatulin1990}:
\begin{subequations}
\label{eq:main_dim_system}
\begin{equation}
\rho_{\tau}+(\rho \tilde{u})_{\xi}=0,
\label{eq:mass_conservation}
\end{equation}
\begin{equation}
\rho\left(\tilde{u}_{\tau}+\tilde{u} \tilde{u}_{\xi}\right)+P_{\xi}=0,\hfill
\label{eq:momentum_conservation}
\end{equation}
\begin{eqnarray}
\rho_{l}\left(R R_{\tau\tau}+\frac{3}{2}R_{\tau}^{2}+\frac{4\nu_{l}}{3R}R_{\tau}\right)-
\frac{\rho_{l}}{c_{l}}\Big(R^{2}R_{\tau\tau\tau}+
6RR_{\tau}R_{\tau\tau}+R_{\tau}^{3}\Big)= \nonumber \\ =P_{g}-P-\frac{2\sigma}{R},
\label{eq:Reley_equation}
\end{eqnarray}
\begin{eqnarray}
P_{g}=P_{g,0}\left(\frac{R_{0}}{R}\right)^{3}\Bigg\{1-\frac{(\gamma_{g}-1)R_{0}^{3}}{5 \gamma_{g}\chi_{g}}\frac{R_{\tau}}{R^{2}}-
\frac{(\gamma_{g}-1)^{2}R_{0}^{6}}{525 \gamma_{g}^{2}\chi_{g}^{2}} \times  \nonumber \\ \times \Bigg[(2+15K_{0}^{'})\frac{R_{\tau}^{2}}{R^{4}}
+\frac{12 \gamma_{g}-7}{3(\gamma_{g}-1)R}\left(\frac{R_{\tau\tau}}{R^{2}}-\frac{2R_{\tau}^{2}}{R^{3}}\right)\Bigg]\Bigg\},\hfill \label{eq:pressure_in_bubble}
\end{eqnarray}
\begin{equation}
\rho=(1-\phi)\,\rho_{l}+\phi\,\rho_{g},
\label{eq:mixture_density}
\end{equation}
\begin{equation}
\phi=V\,\rho, \quad V=\frac{4}{3}\,\pi\,R^{3}\,N. \hfill
\label{eq:gas_concentration}
\end{equation}
\end{subequations}
We use the following notations in system \eqref{eq:main_dim_system}: $\xi$ is cartesian coordinate, $\tau$ is time, $\rho(\xi,\tau)$ is the density of the bubble--liquid mixture, $P(\xi,\tau)$ is the pressure of the mixture, $\tilde{u}(\xi,\tau)$ is the velocity of the mixture, $R=R(\xi,\tau)$  is the bubbles radius, $\rho_{l}, \rho_{g}(\xi,\tau)$ are densities of the liquid and the gas respectively, $P_{g}(\xi,\tau)$ is the pressure of the gas in bubbles, $P_{g,0}$ and $R_{0}$ are the pressure and the radius of bubbles in the unperturbed state, $\phi$ is the volume gas content, $V$ is the specific volume of the gas in the mixture, $\sigma$ is the surface tension, $\nu_{l}$ is the kinematic liquid viscosity, $\gamma_{g}$ is the ratio of the specific heats for the gas, $\chi_{g}$ is the thermal diffusivity of the gas, $K_{g}$ is the thermal conductivity of the gas, $K_{0}^{'}=dK/dT$ at $T=T_{0}$, where $T_{0}$ is the temperature of the mixture in the unperturbed state.

The first two equations from system \eqref{eq:main_dim_system} are the continuity equation and the Euler equation for the mixture. Let us note that at the derivation of equation \eqref{eq:Reley_equation} for bubbles dynamics the liquid viscosity at the inter--phase boundary, the slight liquid compressibility and surface tension were taken into consideration \cite{Keller1980,Prosperetti1986}. Equation \eqref{eq:pressure_in_bubble} was obtained in \cite{Prosperetti1991} under assumption that the gas temperature in the bubble deviates little from the temperature in the unperturbed state. Equations \eqref{eq:mixture_density}, \eqref{eq:gas_concentration} are definitions of the gas--liquid mixture density, the volume gas content and the specific volume of the gas in the mixture correspondingly.

Let us note that the approach based on the theory of thermo--microstretch fluid \cite{Saccomandi1994} can be used for the description of a gas--liquid continuum. For example, acceleration waves in a thermo--microstretch fluid were studied in \cite{Saccomandi1994}.

We suppose that the pressure and density of the mixture in the unperturbed state are constants and all bubbles have the same radius and are uniformly distributed in the liquid.

Assuming that the volume gas content is small  $\phi \ll 1$ from \eqref{eq:mixture_density}---\eqref{eq:gas_concentration} we obtain
\begin{equation}
\rho=\frac{\rho_{l}}{1+\rho_{l}\,V}, \quad V=\frac{4}{3}\,\pi\,R^{3}\,N.
\label{eq: density}
\end{equation}
This equation connects the density of the bubble--liquid mixture with the bubbles radius.

We use the following initial conditions:
\begin{equation*}
   t=0: \quad P=P_{g}=P_{0},\,\,\,\,\,P_{0}=\mbox{const}.
\end{equation*}

Let us suppose that deviation of the mixture density is small:
\begin{equation}
\begin{gathered}
 \rho(\xi,\tau) = \rho_{0} +\delta \tilde{\rho}(\xi,\tau),  \quad
  \rho_{0}=\mbox{const}, \\ \delta=\frac{||\rho-\rho_{0}||}{\rho_{0}}\ll 1,\quad
  \rho(\xi,0)=\rho_{0},
  \label{eq: rel1}
  \end{gathered}
\end{equation}
where $\delta$ is a small parameter corresponding to small deviations of the mixture density from its equilibrium value.

Using formula (\ref{eq: rel1}) from \eqref{eq: density} with accuracy up to $\delta^{2}$ we obtain
\begin{equation}
\begin{gathered}
 R=R_{0}-\chi\delta\tilde{\rho}+\chi_{1}\delta^{2}\tilde{\rho}^{2},\vspace{0.1cm}\\
 R_{0}^{3}=\frac{3}{4\pi N}\left(\frac{1}{\rho_{0}}-\frac{1}{\rho_{l}}\right), \quad \chi=\frac{R_{0}}{3\,\rho_{0}^{2}\,V_{0}}, \quad
\chi_{1}=\frac{R_{0}(3\,\rho_{0}V_{0}-1)}{9\,\rho_{0}^{4}\,V_{0}^{2}}.
  \label{eq:R_equation}
  \end{gathered}
\end{equation}

Substituting \eqref{eq: rel1} and \eqref{eq:R_equation} into equations \eqref{eq:mass_conservation}--\eqref{eq:pressure_in_bubble} and using the dimensionless variables
\begin{eqnarray}
\xi = L\, \xi^{'},  \quad   \tau= \frac{ L }{ c_{0} }\, \tau', \quad \tilde{u} =\delta c_{0}\, \tilde{u}^{'}, \quad
  \tilde{\rho}=\rho_{0} \tilde{\rho}^{'}, \quad P =\delta P_0\, P'+P_0-\frac{2\sigma}{R_{0}},
  \label{eq: non-dim_subst}
\end{eqnarray}
we have the following system of equations (primes are omitted)
\begin{align}
&\tilde{\rho}_{\tau}+\tilde{u}_{\xi}+\delta (\tilde{\rho} \tilde{u})_{\xi}=0, \nonumber \\
&(\delta\tilde{\rho}+1)(\tilde{u}_{\tau}+\delta \tilde{u} \tilde{u}_{\xi})+\frac{1}{\tilde{\alpha}} P_{\xi}=0,\nonumber \\
&P=\tilde{\alpha}\tilde{\rho}+\delta\tilde{\alpha}_{1}\tilde{\rho}^{2}+
\tilde{\beta}\tilde{\rho}_{\tau\tau}-\delta\tilde{\beta}_{1}\tilde{\rho}\,\tilde{\rho}_{\tau\tau}-
\delta\tilde{\beta}_{2}\tilde{\rho}_{\tau}^{2}+\nonumber \\+
&\tilde{\mu}\tilde{\rho}_{\tau}+\delta\tilde{\mu}_{1}\tilde{\rho}\tilde{\rho}_{\tau}
-\tilde{\gamma}\tilde{\rho}_{\tau\tau\tau}+\delta\tilde{\gamma}_{1}\tilde{\rho}\tilde{\rho}_{\tau\tau\tau}+
3\delta\tilde{\gamma}_{1}\tilde{\rho}_{\tau}\tilde{\rho}_{\tau\tau}.
  \label{eq:main_non_dym_system}
\end{align}
Here $L$ and $\tau^{*}=L/c_{0}$ is the characteristic length scale and the characteristic time for our problem correspondingly and $c_{0}$ is the speed of linear waves in the liquid with gas bubbles:
\begin{equation}
c_{0}^{2}=\frac{3\tilde{\mu}P_{0}}{R_{0}}-\frac{2\sigma\tilde{\mu}}{R_{0}^{2}}.
\label{eq:linear_waves_speed}
\end{equation}
Expressions for nondimensional parameters $\tilde{\alpha}$, $\tilde{\beta}$, $\tilde{\alpha}_{1}$, $\tilde{\alpha}_{2}$, $\tilde{\beta}_{1}$, $\tilde{\beta}_{2}$, $\tilde{\mu}$, $\tilde{\mu}_{1}$, $\tilde{\gamma}$, $\tilde{\gamma}_{1}$ are presented in \ref{sec:A}.

In the case of incompressible liquid, system of equations \eqref{eq:main_non_dym_system} is transformed to the system of equations considered in \cite{Kudryashov2013}. If we neglect the liquid compressibility and the interphase heat transfer we obtain the model considered in works \cite{Wijngaarden1968,Nakoryakov1972,Wijngaarden1972,Nakoryakov1993,Nigmatulin1990}.

System of equations \eqref{eq:main_non_dym_system} consists of two conservation laws (corresponding to conservation of mass and  momentum) and the dynamical state equation for the liquid containing gas bubbles. The complexity of \eqref{eq:main_non_dym_system} is connected with high order derivatives and nonlinear terms in the dynamical state equation.

Systems similar to \eqref{eq:main_non_dym_system} can be used for studying nonlinear waves in plasma physics, fluid dynamics, long waves in ferromagnetic media, nonlinear optics etc. \cite{Su1969,Leblond2008}. Thus, one can consider system of equations \eqref{eq:main_non_dym_system} as a prototypical system for the description of nonlinear waves in a nonlinear media with dispersion and dissipation. It is worth noting that there are two factors of waves damping in system \eqref{eq:main_non_dym_system}: the first one mainly impacts on the damping of long waves and the second one mainly impacts on the damping of short waves. Linear terms with the first and third order derivatives in the dynamical state equation correspond to these damping factors.

The dispersion relation for the linearization of \eqref{eq:main_non_dym_system} has the form
\begin{equation}
k^{2}=\frac{\omega^{2}}{1-(\tilde{\beta}/\tilde{\alpha})\omega^{2}+
i\left[(\tilde{\mu}/\tilde{\alpha})\omega+(\tilde{\gamma}/\tilde{\alpha})\omega^{3}\right]}.
\label{eq:main_dispersion_relation}
\end{equation}
We see that system of equations \eqref{eq:main_non_dym_system} in the linear case describes superposition of two waves propagating in opposite directions with the phase speed $\mbox{Re}(\omega/k)$ and the damping factor $\mbox{Im}(\omega/k)$. Using values of parameters \eqref{eq: non-dim_parameters} one can obtain that the dispersion of waves is determined by the inertia of gas bubbles and the interphase heat transfer. Analogously, we see that the dissipation of waves is determined by the liquid viscosity, liquid compressibility and interphase heat transfer.

Now we consider long weakly nonlinear waves governed by system of equations \eqref{eq:main_non_dym_system}. We assume that the characteristic wavelength is much greater than the equilibrium radius of bubbles. Thus, we have the small parameter $\varepsilon$ equal to the ratio of bubbles radius in the unperturbed state $R_{0}$ to the characteristic wavelength $L$ ($\varepsilon=R_{0}/L \ll 1$). Assuming that $\omega$ in \eqref{eq:main_dispersion_relation} is small we obtain
\begin{equation}
k=-\omega+\frac{i\tilde{\mu}}{2\tilde{\alpha}}\omega^{2}-
\left(\frac{\tilde{\beta}}{2\tilde{\alpha}}-\frac{3\tilde{\mu}^{2}}{8\tilde{\alpha}^{2}}\right)\omega^{3}+\ldots \, .
\label{eq:main_dispersion_relation_expansion}
\end{equation}
From \eqref{eq:main_dispersion_relation_expansion} we see that system of equations \eqref{eq:main_non_dym_system} contains at least two different time and length scales: that of the main influence of dissipation and that of the main influence of dispersion. Below we consider nonlinear waves on these two scales and obtain governing nonlinear differential equations. We also assume that $\delta$ has the same order as $\varepsilon$ ($\delta \sim \varepsilon$).

For the derivation of the equations for nonlinear waves we use the reductive perturbation method \cite{Washimi1966,Su1969,Leblond2008}. We introduce the 'slow' variables
\begin{equation}
\begin{gathered}
  x = \varepsilon^m(\xi-\tau), \quad t= \varepsilon^{m+1}\, \tau ,\quad
  m > 0, \quad \varepsilon \ll 1,
  \label{eq: rescaling_coordinates}
    \end{gathered}
\end{equation}
and search for the solutions of \eqref{eq:main_non_dym_system} in the form of a series in the small parameter $\varepsilon$
\begin{equation}
  \begin{gathered}
    \tilde{\rho}  = \rho_1 + \varepsilon \rho_2 +\ldots,  \quad
    \tilde{u} =  u_1 + \varepsilon u_2 +  \ldots,  \quad
    P =  P_1  + \varepsilon P_2   +\ldots\, .
  \end{gathered}
  \label{eq: asymptotic_expansion1}
\end{equation}
We introduce the parameter $m$ in variables \eqref{eq: rescaling_coordinates} in order to take into account two different time and length scales of waves in the bubble--liquid mixture. Varying the parameter $m$ we obtain nonlinear differential equations governing nonlinear waves on corresponding time and length scales. Also the main influence of the different physical properties (e.g. dissipation, dispersion) corresponds to different values of $m$. In this way we can reveal main physical properties affecting nonlinear waves at the corresponding length and time scales.

Substituting \eqref{eq: rescaling_coordinates}, \eqref{eq: asymptotic_expansion1} into \eqref{eq:main_non_dym_system} and collecting coefficients at $\varepsilon^{0}$ we find
\begin{equation}
  \begin{gathered}
u_{1}(x,t)=\rho_{1}(x,t), \quad P_{1}(x,t)=\tilde{\alpha} \rho_{1}(x,t).
  \end{gathered}
  \label{eq: first_order_relation}
\end{equation}
Collecting coefficients at $\varepsilon$ and using relations \eqref{eq: first_order_relation} we obtain the equation
\begin{equation}
  \begin{gathered}
\rho_{1t}+\left(1+\frac{\tilde{\alpha}_{1}}{\tilde{\alpha}}\right)\rho_{1}\rho_{1x}+
\varepsilon^{2m-1}\frac{\tilde{\beta}}{2\tilde{\alpha}}(\rho_{1xxx}-2\varepsilon\rho_{1txx})-\hfill\\
-\varepsilon^{2m}\frac{\tilde{\beta}_{1}}{2\tilde{\alpha}}\left(\rho_{1x}\rho_{1xx}+\rho_{1}\rho_{1xxx}\right)
-\varepsilon^{2m}\frac{\tilde{\beta}_{2}}{\tilde{\alpha}}\rho_{1x}\rho_{1xx}+ \hfill\\
+\varepsilon^{3m-1}\frac{\tilde{\gamma}}{2\tilde{\alpha}}
\rho_{1xxxx}=
\varepsilon^{m-1} \frac{\tilde{\mu}}{2\tilde{\alpha}}(\rho_{1xx}-\varepsilon \rho_{1tx})
+\varepsilon^{m} \frac{\tilde{\mu}_{1}}{2\tilde{\alpha}} (\rho_{1}\rho_{1x})_{x}. \hfill
\label{eq: main_evolution_equation}
  \end{gathered}
\end{equation}
We see that right--going nonlinear waves are governed by equation \eqref{eq: main_evolution_equation}. Below we consider two different cases of \eqref{eq: main_evolution_equation}. The first one is the case of the main influence of dissipation  ($m=1$). The other one is the case of the main influence of dispersion ($m=1/2$). Thus, we see that dissipation mainly influences waves propagation at the longer length and time scales than the dispersion. We drop in \eqref{eq: main_evolution_equation} three terms of order $\varepsilon^{3m}$ because they will not appear in our further consideration.

Later we will take into account in \eqref{eq: main_evolution_equation} only terms to the first order of $\varepsilon$. Thus, we consider the terms of the lowest order ($\varepsilon^{0}$) and the next one ($\varepsilon$). At the lowest order in $\varepsilon$, one can obtain known equations for the waves in liquid with gas bubbles that is the Burgers equation and the Burgers--Korteweg--de Vries equation. At the next order we find new equations for more accurate description of nonlinear waves. Also taking into consideration high order terms with respect to the small parameter we reveal new  physical effects affecting nonlinear waves propagation.

\section{The case of $m=1$: main influence of dissipation on nonlinear waves propagation}
Let us consider time and length scales where dissipation mainly influences nonlinear waves. Substituting $m=1$ in \eqref{eq: main_evolution_equation} and taking into account terms with $\varepsilon$ in zero and first powers we obtain
\begin{eqnarray}
\rho_{1t}+\left(1+\frac{\tilde{\alpha}_{1}}{\tilde{\alpha}}\right)\rho_{1}\rho_{1x}-\frac{\tilde{\mu}}{2\tilde{\alpha}}\rho_{1xx}
=\varepsilon\left(\frac{\tilde{\mu}_{1}}{2\tilde{\alpha}} (\rho_{1}\rho_{1x})_{x}-\frac{\tilde{\beta}}{2\tilde{\alpha}}(\rho_{1xxx}) -\frac{\tilde{\mu}}{2\tilde{\alpha}}\rho_{1tx}\right).
\label{eq:ext_Burgers}
\end{eqnarray}
We see that \eqref{eq:ext_Burgers} is the generalization of the Burgers equation \cite{Wijngaarden1972,Nakoryakov1993,Nigmatulin1990} for long weakly nonlinear waves in the liquid with gas bubbles. Equation \eqref{eq:ext_Burgers} was derived for the first time in \cite{Kudryashov2013} for the description of waves in the liquid containing gas bubbles.

\begin{figure}
\center
 \includegraphics[width=0.55\textwidth]{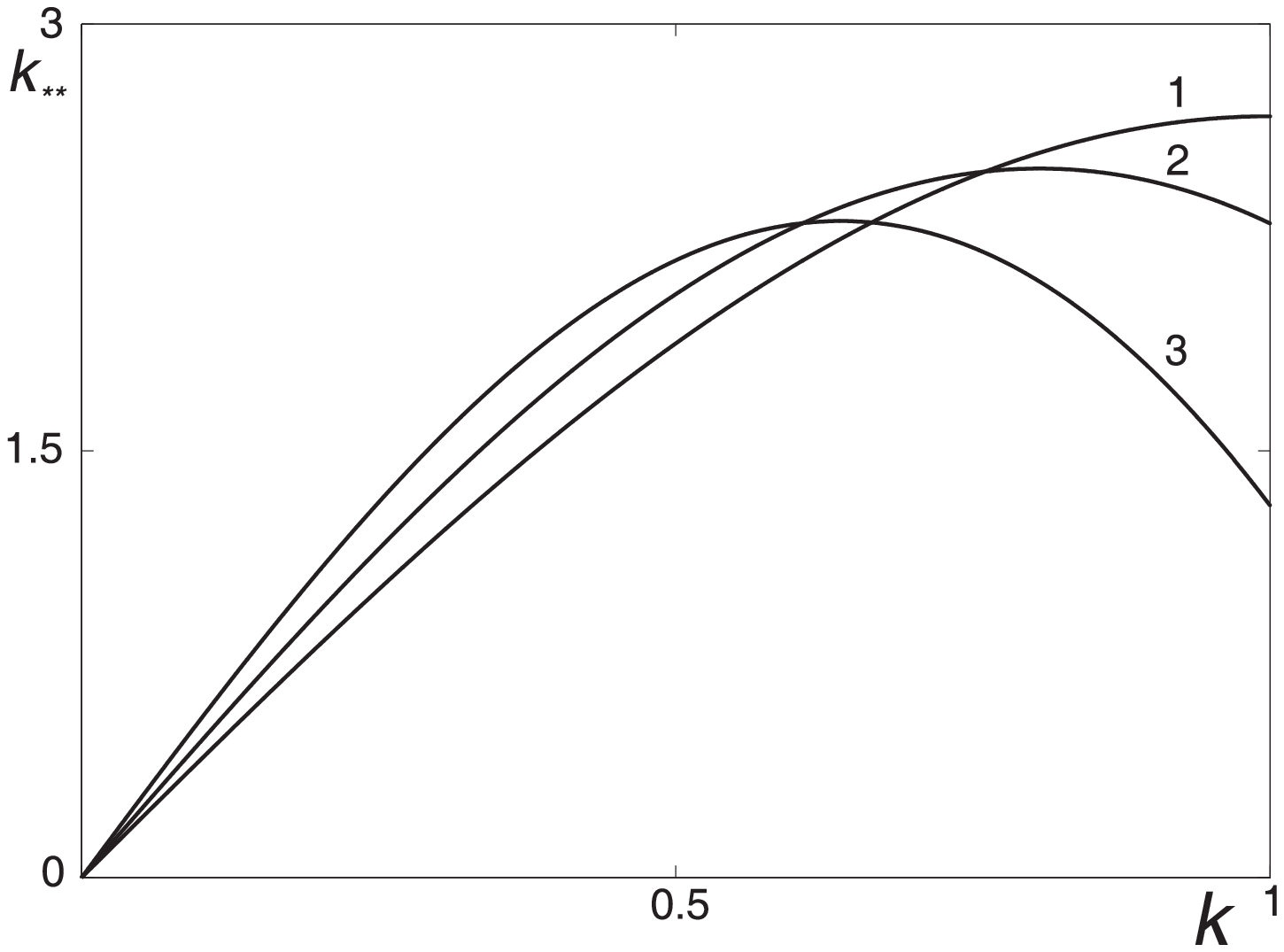} 
    \caption{Variation of the linear damping coefficient with the wave number $k$ for  waves in water containing carbon dioxide bubbles governed by \eqref{eq:ext_Burgers_1} for different values of the equilibrium bubbles radius  $R_{0}=0.23$mm (Curve 1), $R_{0}=0.25$mm (Curve 2) and  $R_{0}=0.27$mm (Curve 3) with  $\phi_{0}=0.001$.}
  \label{fig:1.1}
\end{figure}

Let us use the following notations in  \eqref{eq:ext_Burgers}:
\begin{align}
& \rho_{1}=u, \quad \left(1+\frac{\tilde{\alpha}_{1}}{\tilde{\alpha}}\right)=\alpha, \quad \frac{\tilde{\beta}}{2\tilde{\alpha}}=\beta, \quad,
 \frac{\tilde{\mu}}{2\tilde{\alpha}}=\mu, \quad \frac{\tilde{\mu}_{1}}{2\tilde{\alpha}}=\nu,
\label{eq:ext_Burgers_notations}
\end{align}
then we have
\begin{equation}
u_{t}+\alpha u u_{x}-\mu u_{xx}=\varepsilon\bigg(\nu (uu_{x})_{x}-\beta u_{xxx}-\mu u_{xt}\bigg).
\label{eq:ext_Burgers_1}
\end{equation}

Linearizing \eqref{eq:ext_Burgers_1} under the trivial solution we find corresponding dispersion relation
\begin{equation}
\begin{gathered}
 \omega=\frac{i \varepsilon \beta k^{3}-\mu k^{2} }{i-\varepsilon \mu k}
=\frac{1}{1+\varepsilon^{2}\mu^{2}k^{2}}\Bigg[\varepsilon(\mu^{2}+\beta)k^{3}+i\mu k^{2}(1-\varepsilon^{2}\beta k^{2})\Bigg]\simeq  \\
 \simeq \varepsilon(\mu^{2}+\beta)k^{3}+i\mu k^{2}.
\label{eq:ext_Burgers_dispersion_relation}
\end{gathered}
\end{equation}
From \eqref{eq:ext_Burgers_dispersion_relation} we see that waves attenuate with the linear damping coefficient proportional to the parameter $\mu$. There is also weak dispersion of waves proportional to $\mu^{2}+\beta$. Using \eqref{eq: non-dim_parameters} and \eqref{eq:ext_Burgers_notations} we find that dissipation of waves governed by \eqref{eq:ext_Burgers_1} is determined by the liquid viscosity and thermal diffusivity of the gas. Dispersion of waves governed by \eqref{eq:ext_Burgers_1} depends on the radius of bubbles in the unperturbed state, the volume gas content in the unperturbed state, the liquid viscosity and thermal diffusivity of the gas. We see that the phase and the group speed of nonlinear waves is the result of the interaction between the dispersion and the dissipation. One can take into account this phenomenon only if we consider high order corrections to the Burgers equation. Let us also note that according to dispersion relation \eqref{eq:ext_Burgers_dispersion_relation} the phase speed, the group speed and the linear damping coefficient are bounded for the all values of $k$.

Let us consider the dependence of the linear damping coefficient ($k_{**}=\mbox{Im}(\omega/k)$) on the wave number at different values of equilibrium bubble radius $R_{0}$ for water with bubbles of carbon dioxide. From Fig.\ref{fig:1.1} we see that the linear damping coefficient decreases when the equilibrium bubbles radius increases. This can be treated as the increase of the dispersion impact with the increase of bubbles radius. We also see from Fig.\ref{fig:1.1} that the attenuation of short waves for smaller values of $R_{0}$ is more intensive than the attenuation of long waves. It is worth noting that the main contribution to the damping parameter $\mu$ is given by the interphase heat transfer. The dependence of the phase speed ($v_{\mbox{\tiny{phase}}}=\mbox{Re}(\omega/k)$) on the wave number is very similar to the dependence of the linear damping coefficient on the wave number.

Equation \eqref{eq:ext_Burgers_1} is a prototypical equation at the order of $\varepsilon$ for the nonlinear waves in media with dissipation and weak dispersion. However, for further investigation of \eqref{eq:ext_Burgers_1}  it is better to convert it to the evolution form and to construct its normal form using the near--identity transformations. Applying the near--identity transformations allow us to construct an equation which is asymptotically equivalent to \eqref{eq:ext_Burgers_1} and contains two arbitrary parameters. Taking into account arbitrariness of these parameters we show below that one can use an integrable evolution equation for the description of nonlinear waves in the liquid with gas bubbles. Even though the equation is integrable under a certain condition on physical parameters it can be used for description of waves because this condition is satisfied for real liquids with gas bubbles. In the general case (without imposing any conditions on physical parameters) one can use arbitrariness of the parameters introduced by the near--identity transformations to obtain exact solutions in a more simple and convenient form.

We differentiate \eqref{eq:ext_Burgers_1} once with respect to $x$ and substitute the result into \eqref{eq:ext_Burgers_1}. Then, to the first order in $\varepsilon$, we obtain
\begin{equation}
u_{t}+\alpha u u_{x}-\mu u_{xx}=\varepsilon\bigg((\nu+\mu\alpha) (uu_{x})_{x}-(\beta+\mu^{2}) u_{xxx}\bigg).
\label{eq:ext_Burgers_2}
\end{equation}
We see that \eqref{eq:ext_Burgers_2} is an evolution equation. Now using near--identity transformations we obtain a family of asymptotically equivalent equations for the description of nonlinear waves in the liquid with gas bubbles. Substituting \cite{Kraenkel1998,Veksler2005}
\begin{equation}
u=v+\varepsilon (\lambda_{1} v^{2}+\lambda_{2} v_{x}\partial_{x}^{-1}v),
\label{eq:Kodama_transformations_B}
\end{equation}
into \eqref{eq:ext_Burgers_2}, to the first order in $\varepsilon$, we have the equation
\begin{equation}
\begin{gathered}
v_{t}+\alpha v v_{x}-\mu v_{xx}=\varepsilon\Bigg((2\mu\lambda_{2}+\mu\alpha+\nu)v v_{xx}+(2\mu\lambda_{1}+\mu\alpha+\nu)v_{x}^{2}-\hfill \\-\frac{\alpha(\lambda_{2}+2\lambda_{1})}{2}v^{2}v_{x}
-(\beta+\mu^{2}) v_{xxx}\Bigg).
\label{eq:ext_Burgers_3}
 \end{gathered}
\end{equation}
Here $\lambda_{1},\lambda_{2}$ are arbitrary parameters of Kodama transformations \eqref{eq:Kodama_transformations_B}. Choosing various values of parameters $\lambda_{1},\lambda_{2}$ one can obtain different nonlinear evolution equations for long weakly nonlinear waves in the liquid containing gas bubbles. Thus, equation \eqref{eq:ext_Burgers_3} is a family of asymptotically equivalent equations for nonlinear waves at the quadratic order in the case of dissipation main influence. Let us note that equation \eqref{eq:ext_Burgers_3}  was considered in works \cite{Kraenkel1998,Veksler2005} where applications of this equation to the acoustic waves and to the flow of viscous gas were considered.

It is worth noting that the last equality from \eqref{eq:ext_Burgers_dispersion_relation} is exactly the dispersion relation for \eqref{eq:ext_Burgers_3}. Thus, transformations \eqref{eq:Kodama_transformations_B} with the procedure of the excluding mixed derivative do not change the dispersion relation to the first order in $\varepsilon$ as it was pointed out in \cite{Dullin2003,Dullin2004}.

\begin{figure}
\center
 \includegraphics[width=0.55\textwidth]{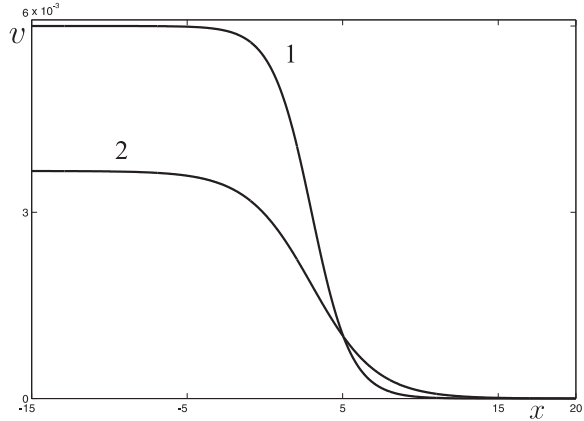}
  \caption{The one--kink solution of the Burgers equation (curve 1) and the one--kink solution of equation \eqref{eq:ext_Burgers_3} under conditions \eqref{eq:STO_parameters3} (curve 2) for water containing carbon dioxide bubbles with $\phi_{0}=0.001$ and $R_{0}=3.4 \cdot 10^{-3}$mm.  }
  \label{fig:1a}
\end{figure}

One can find that \eqref{eq:ext_Burgers_3} is integrable under certain conditions on parameters $\lambda_{1}$, $\lambda_{2}$ and $\nu$.  Indeed, let us consider the following values of parameters $\lambda_{1},\lambda_{2}$:
\begin{subequations}
\label{eq:STO_parameters3}
\begin{equation}
\lambda_{1}=\lambda_{2}=\frac{(\beta+\mu^{2})\alpha}{2\mu^{2}},\\
\label{eq:STO_parameters}
\end{equation}
\begin{equation}
\nu=\frac{\alpha(\beta-\mu^{2})}{2\mu}. \hfill
\label{eq:STO_parameters1}
\end{equation}
\end{subequations}
Then from \eqref{eq:ext_Burgers_3} after applying the following transformations
\begin{equation}
\begin{gathered}
x=\frac{3(\beta+\mu^{2})\varepsilon}{\mu}\,\bigg(x^{'}-3t^{'}\bigg),\;
t=\left(\frac{3}{\mu}\right)^{3}(\beta+\mu^{2})^{2}\,\varepsilon^{2}\,t^{'},\;
v=\frac{2\mu^{2}}{3\alpha(\beta+\mu^{2})\varepsilon}(v^{'}-1),
\label{eq: STO1_tr}
 \end{gathered}
\end{equation}
we obtain the equation (primes are omitted)
\begin{equation}
v_{t}+v_{xxx}-3(v\,v_{x})_{x}+3v^{2}v_{x}=0.
\label{eq:ext_Burgers_5}
\end{equation}
Equation \eqref{eq:ext_Burgers_5} is the second member of the Burgers hierarchy and is often called the Sharma--Tasso--Olver equation \cite{Tasso1996,Olver1977,Kudryashov2009,Kudryashov2012a}. This equation can be linearized by the Cole--Hopf transformation. Various exact solutions of \eqref{eq:ext_Burgers_5} were obtained in \cite{Kudryashov2009,Kudryashov2012a}.

Equation \eqref{eq:ext_Burgers_5}  was obtained under condition \eqref{eq:STO_parameters1} on the physical parameter $\nu$. Using \eqref{eq: non-dim_parameters} we find that condition \eqref{eq:STO_parameters1} can be realized for the real liquid with gas bubbles. Let us note that an analogous condition for \eqref{eq:ext_Burgers_3} in the case of waves in viscous gas leads to nonphysical values of parameters \cite{Kraenkel1998}.

\begin{table}[!h]
\center
\caption{\label{tab:table1}Values of bubbles radius $R_{0}$ in the unperturbed state obtained from condition \eqref{eq:STO_parameters1}.}
\begin{tabular}{lc}   \toprule
Carrier phase& Radius in the unperturbed state $R_{0}$,mm\\ \midrule
Water & $3.4 \cdot 10^{-3}$ \\ 
Oil & $0.35$ \\  
Glycerine & $5.1$ \\ 
\bottomrule
\end{tabular}
\end{table}

Let us analyze condition \eqref{eq:STO_parameters1} in more detail. To simplify calculations we neglect the influence of the interphase heat transfer. Using \eqref{eq: non-dim_parameters} from \eqref{eq:STO_parameters1} we obtain an equation for the equilibrium bubbles radius $R_{0}$. Solving this equation for different carrier phases we obtain values of $R_{0}$ that are listed in Tab.\ref{tab:table1}. From Tab.\ref{tab:table1} one can see that these values of $R_{0}$ correspond to real liquids with gas bubbles.

It is interesting to compare the one--kink solution of equation \eqref{eq:ext_Burgers_3} under conditions \eqref{eq:STO_parameters3} with the one--kink solution of the Burgers equation. Plots of these solutions at the time moment $t=1$ are presented in Fig.\ref{fig:1a}. From Fig.\ref{fig:1a} we see that the wave governed by the perturbed Burgers equation has a larger width of the wave front than the  wave governed by the Burgers equation. This fact can be explained by the presence of additional nonlinear dissipative terms in \eqref{eq:ext_Burgers_3}. We can take into account these terms only if we consider high order corrections to the Burgers equations. We also can see that dissipation connected with these terms is caused by both the interphase heat transfer and the liquid viscosity.

Let us construct solutions of \eqref{eq:ext_Burgers_3} without imposing condition \eqref{eq:STO_parameters1}. In this case equation \eqref{eq:ext_Burgers_3} is not integrable. Thus, we have to choose parameters $\lambda_{1},\lambda_{2}$ in such a way that the Laurent expansion of the general solution has a simple form.

Let
\begin{equation}
\lambda_{1}=\lambda_{2}=-\frac{2}{4\mu+\alpha}(2\beta+2\mu^{2}+\mu\alpha+\nu),
\end{equation}
then equation \eqref{eq:ext_Burgers_3} has the following traveling wave solution
\begin{equation}
\begin{gathered}
v=\frac{(2\mu+\alpha)(4\mu+\alpha)}
{4\varepsilon(3\beta\alpha+3\mu^{2}\alpha+\mu\alpha^{2}+\nu\alpha+4\mu\beta+4\mu^{3})}
+\sqrt{B}\tanh\{\sqrt{B}(z-z_{0})\}, \\ z=x-C_{0}t,
\label{eq:ext_Burgers_3_solution}
\end{gathered}
\end{equation}
where $C_{0}$ is an arbitrary traveling wave speed, $z_{0}$ is an arbitrary constant and $B$ is the parameter defining the wave amplitude and the width of the wave front. The parameter $B$ depends on physical parameters $\mu,\nu,\alpha,\beta$ and the traveling wave speed $C_{0}$.
Explicit expression for $B$ is presented in \ref{sec:B}. Solution \eqref{eq:ext_Burgers_3_solution} can be obtained applying one of the ansatz methods for finding exact solutions. We used the simplest equation method \cite{Kudryashov2005}.



One can see that solution \eqref{eq:ext_Burgers_3_solution} is a weak shock wave. Using values of parameters $\mu,\nu,\alpha,\beta$ presented in \ref{sec:A} one can find that the width of the wave front of \eqref{eq:ext_Burgers_3_solution} increases with the bubbles radius in the unperturbed state ($R_{0}$). The amplitude of solution \eqref{eq:ext_Burgers_3_solution} is inversely related to $R_{0}$.

\section{The case of $m=1/2$: main influence of dispersion on the nonlinear waves propagation}

Let us consider the length and time scales where dispersion mainly impacts on nonlinear waves propagation. In this case from \eqref{eq: main_evolution_equation} we obtain the generalization of the Burgers--Korteweg--de Vries equation.

Indeed, substituting $m=1/2$ into \eqref{eq: main_evolution_equation}, to the first order in $\varepsilon$, we obtain
\begin{equation}
\begin{gathered}
u_{t}+\alpha u u_{x}+\beta u_{xxx} -\mu u_{xx}- \varepsilon\Big(2 \beta u_{txx} + \beta_{1}u u_{xxx} + \beta_{2} u_{x}u_{xx}  - \mu u_{xt}+\\+\nu (u u_{x})_{x}-\gamma u_{xxxx}\Big)=0,
\label{eq:extended_BKdV_equation_1}
 \end{gathered}
\end{equation}
where the following notations were used
\begin{equation}
\begin{gathered}
\rho_{1}=u, \;
\left(1+\frac{\tilde{\alpha}_{1}}{\tilde{\alpha}}\right)=\alpha, \; \frac{\tilde{\beta}}{2\tilde{\alpha}}=\beta,\;
\frac{\tilde{\beta}_{1}}{2\tilde{\alpha}}=\beta_{1},\; \frac{\tilde{\beta}_{1}+2\tilde{\beta}_{2}}{\tilde{\alpha}}=\beta_{2}, \;
\frac{\tilde{\mu}}{2\tilde{\alpha}}=\varepsilon^{1/2}\mu,\vspace{0.1cm} \\ \frac{\tilde{\mu}_{1}}{2\tilde{\alpha}}=\varepsilon^{1/2}\nu,\;
\frac{\tilde{\gamma}}{2\tilde{\alpha}}=\varepsilon^{1/2}\gamma.
\label{eq:extended_BKdV_notations}
 \end{gathered}
\end{equation}
To the best of our knowledge equation \eqref{eq:extended_BKdV_equation_1} is obtained for the first time. We see that equation \eqref{eq:extended_BKdV_equation_1} at $\varepsilon\rightarrow 0$ transforms the Burgers--Korteweg--de Vries equation. Particular cases of \eqref{eq:extended_BKdV_equation_1} were considered in \cite{Kudryashov2010} for waves in the gas--liquid mixture. In the purely dispersive case (i.e. $\mu=\nu=\gamma=0$) equation \eqref{eq:extended_BKdV_equation_1} is the generalization of the Korteweg--de Vries equation \cite{Dullin2003,Dullin2004}. Exact solutions for particular cases of this equation were considered in \cite{Biswas2010,Randruut2011,Randruut2014}.

\begin{figure}
\center
\includegraphics[width=0.55\textwidth]{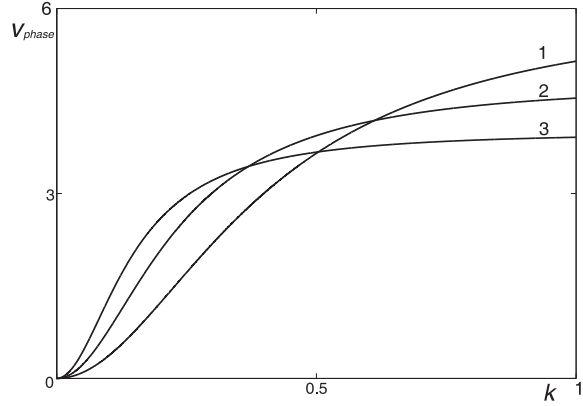}
\caption{Variation of the phase speed with the wave number $k$ for waves in water containing carbon dioxide bubbles governed by \eqref{eq:extended_BKdV_equation_1} for different values of the equilibrium bubbles radius $R_{0}=0,16$mm (Curve 1), $R_{0}=0.2$mm (Curve 2) and $R_{0}=0.25$mm (Curve 3) with $\phi_{0}=0.001$. }
  \label{fig:2b}
\end{figure}

The dispersion relation for \eqref{eq:extended_BKdV_equation_1} has the form
\begin{equation}
\begin{gathered}
\omega=\frac{i\beta k^{3}-\mu k^{2}-\varepsilon \gamma k^{4} }{i(1+2\beta k^{2}\varepsilon)-\varepsilon\mu k}=\vspace{0.1cm}\\=
\frac{k^{3}[\beta+\varepsilon(2k^{2}\beta^{2}+\mu^{2})+\varepsilon^{2}\gamma\mu k^{2}]+ik^{2}[\mu+\varepsilon(\mu\beta+\gamma)k^{2}+2\varepsilon^{2}\gamma\beta k^{4}]}{1+\varepsilon k^{2}[4\beta+\varepsilon(4k^{2}\beta^{2}+\mu^{2})]}\simeq\vspace{0.1cm}\\
\simeq \beta k^{3}+\varepsilon (\mu^{2}-2\beta^{2}k^{2})k^{3}+i (\mu k^{2}+\varepsilon (\gamma-3\beta\mu) k^{4}).
\label{eq:extended_BKdV_equation_dispersion_relation}
 \end{gathered}
\end{equation}

From \eqref{eq:extended_BKdV_equation_dispersion_relation} we see that there are several mechanisms of waves dispersion and attenuation. Dispersion of waves depends on the wavelength. One can see strong dispersion of long waves and weak dispersion of short waves.
Waves dispersion is governed by parameters $\beta$ and $\mu$. We can see two mechanisms of waves attenuation as well: strong attenuation of long waves and weak attenuation of short waves. Parameters $\mu$, $\gamma$ and $\beta$ define waves damping.  Using \eqref{eq: non-dim_parameters} one can see that waves dispersion is caused by the presence of bubbles and by the interphase heat transfer. Waves dissipation is connected with the liquid viscosity, interphase heat transfer and liquid compressibility.

From Fig.\ref{fig:2b} we see that when the radius of bubbles in the unperturbed state increases the phase speed decreases. This is due to the reducing of the heat transfer impact on nonlinear waves propagation. The dependence of the linear damping coefficient on the wave number is similar to the dependence of the phase speed on the wave number.

Let us construct a family of equations asymptotically equivalent to  \eqref{eq:extended_BKdV_equation_1}. In this way with the help of near--identity transformations we will obtain a nonlinear evolution equation for waves in the gas--liquid mixture that contains three arbitrary parameters. Taking into consideration arbitrariness of these parameters one can find that in the case of negligible dissipation nonlinear waves are governed by one of the integrable fifth order evolution equations. In the general case one can use arbitrariness of these parameters for simplifying calculations during construction of exact solutions.

Excluding mixed derivative by the analogy with the previous case and applying near--identity transformations \cite{Kodama1985,Fokas1996,Dullin2003,Dullin2004}:
\begin{equation}
u=v+\varepsilon (\lambda_{1} v^{2}+\lambda_{2} v_{x}\partial_{x}^{-1}v)+\varepsilon\lambda_{3} v_{xx},
\label{eq:Kodama_transformations2}
\end{equation}
from \eqref{eq:extended_BKdV_equation_1} we obtain
\begin{equation}
\begin{gathered}
v_{t}+ \alpha v v_{x} +\beta v_{xxx}-\mu v_{xx} +\varepsilon \Bigg[(6\beta\alpha-\beta_{2}+3\beta\lambda_{2}+6\beta\lambda_{1}-2\alpha\lambda_{3})v_{x}v_{xx}+\hfill \\
+(2\beta\alpha-\beta_{1}+3\beta\lambda_{2})v v_{xxx}
+\frac{\alpha(2\lambda_{1}+\lambda_{2})}{2}v^{2}v_{x}-(2\mu\lambda_{1}+\mu\alpha+\nu)v_{x}^{2}-\hfill \\-(2\mu\lambda_{2}+\mu\alpha+\nu)vv_{xx}
+\mu^{2} v_{xxx}+(\gamma-3\beta\mu)v_{xxxx}+2\beta^{2} v_{xxxxx}\Bigg]=0,\hfill
\label{eq:extended_BKdV_equation_3}
 \end{gathered}
\end{equation}
where $\lambda_{1},\lambda_{2},\lambda_{3}$ are arbitrary parameters.
Equation \eqref{eq:extended_BKdV_equation_3} is a family of the generalizations of the Burgers--Korteweg--de Vries equation and is obtained for the first time. Let us note that the dispersion relation for \eqref{eq:extended_BKdV_equation_3}  is equal to the right--hand side of \eqref{eq:extended_BKdV_equation_dispersion_relation}. Thus, we again see that the Kodama transformations do not change dispersion relations to the first order in $\varepsilon$ \cite{Dullin2003,Dullin2004}.

Below we consider two different cases of \eqref{eq:extended_BKdV_equation_3}. The first one is a purely dispersive case ($\mu=\nu=\gamma=0$). The second one is the general case of \eqref{eq:extended_BKdV_equation_3}.

In the purely dispersive case ($\mu=\nu=\gamma=0$) from \eqref{eq:extended_BKdV_equation_3} we obtain the generalized Korteweg--de Vries equation
\cite{Olver1984,Zhi1997,Fokas1996,Dullin2003,Dullin2004}:
\begin{equation}
\begin{gathered}
v_{t}+ \alpha v v_{x} +\beta v_{xxx} +\varepsilon \Bigg[\frac{\alpha(2\lambda_{1}+\lambda_{2})}{2}v^{2}v_{x}+(2\beta\alpha-\beta_{1}+3\beta\lambda_{2})v v_{xxx}+\hfill\\+2\beta^{2} v_{xxxxx}+(6\beta\alpha-\beta_{2}+3\beta\lambda_{2}+6\beta\lambda_{1}-2\alpha\lambda_{3})v_{x}v_{xx}
\Bigg]=0. \hfill
\label{eq:extended_KdV_equation}
 \end{gathered}
\end{equation}
Equation \eqref{eq:extended_KdV_equation} was well studied, for example exact solutions of \eqref{eq:extended_KdV_equation} were investigated in \cite{Kudryashov2001,Tan2002,Kudryashov2012b}.

The normal form equation for \eqref{eq:extended_KdV_equation} is not unique and depends on values of parameters $\lambda_{1},\lambda_{2},\lambda_{3}$. It can be one of the integrable fifth order evolution equations: the Lax equation, the Kaup--Kupershmidt equation or the Sawada--Kotera equation. The Cauchy problem for these equation can be solved with the help of the inverse scattering transform \cite{Sawada1974,Kupershmidt1980,Lax1968,kaup1980inverse}.

Indeed, let
\begin{equation}
\begin{gathered}
\lambda_{1}=\frac{26\alpha\beta-3\beta_{1}}{18\beta},\; \lambda_{2}=\frac{4\alpha\beta+3\beta_{1}}{9\beta}, \; \lambda_{3}=\frac{28\alpha\beta-3\beta_{2}}{6\alpha},
\label{eq:KdV5_parameters}
 \end{gathered}
\end{equation}
then from \eqref{eq:extended_KdV_equation} we obtain the Lax equation
\begin{equation}
\begin{gathered}
v_{t}+ \alpha v v_{x} +\beta v_{xxx} +2\beta\varepsilon \Bigg[\frac{5\alpha^{2}}{6\beta}v^{2}v_{x}+\frac{5\alpha}{3}[v v_{xxx}+2v_{x}v_{xx}]
+\beta v_{xxxxx}\Bigg]=0.
\label{eq:KdV5}
 \end{gathered}
\end{equation}
By the analogy at
\begin{equation*}
\begin{gathered}
\lambda_{1}=\frac{12\alpha\beta-\beta_{1}}{6\beta},\; \lambda_{2}=\frac{6\alpha\beta+2\beta_{1}}{6\beta}, \; \lambda_{3}=\frac{16\alpha\beta-\beta_{2}}{2\alpha},
 \end{gathered}
\end{equation*}
or at
\begin{equation*}
\begin{gathered}
\lambda_{1}=\frac{12\alpha\beta-\beta_{1}}{6\beta},\; \lambda_{2}=\frac{6\alpha\beta+2\beta_{1}}{6\beta}, \; \lambda_{3}=\frac{34\alpha\beta-4\beta_{2}}{8\alpha},
 \end{gathered}
\end{equation*}
from \eqref{eq:extended_KdV_equation} we obtain the Sawada--Kotera equation or the Kaup--Kupershmidt equation correspondingly.

The one--soliton solution of the Lax equation has the form
\begin{equation}
v(x,t)=v_{\mbox{\tiny{m}}}\cosh^{-2}\left[(x-C_{0}\,t+\phi_{0})/\Delta\right],
\label{eq:KdV5_soliton_solution}
\end{equation}
where $C_{0}$ and $\phi_{0}$ are arbitrary constants and the amplitude $v_{\mbox{\tiny{m}}}$ and the width $\Delta$ are defined by
\begin{equation}
\begin{gathered}
v_{\mbox{\tiny{m}}}=\frac{6\left[1-\sqrt{1+8\varepsilon C_{0}}+5\varepsilon C_{0}\right]}{\alpha\varepsilon(5\sqrt{1+8\varepsilon C_{0}}-3)}, \quad
\Delta=\frac{4\beta\varepsilon}{[\beta\varepsilon(\sqrt{1+8\varepsilon C_{0}}-1)]^{1/2}}. \hfill
\label{eq:KdV5_solitary_wave_speed}
\end{gathered}
\end{equation}

\begin{figure}
\center
 \includegraphics[width=0.55\textwidth]{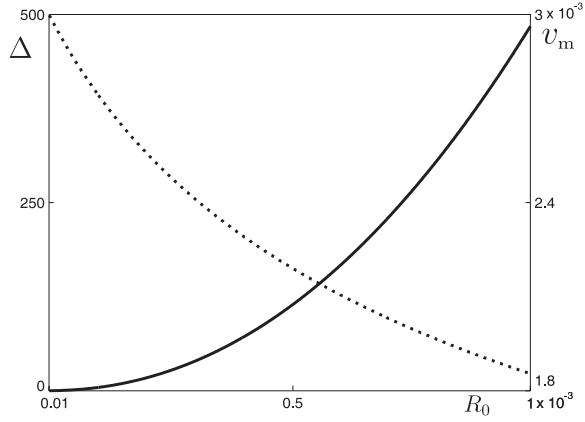}
     \caption{Variation of the width (shown by solid curve) and the amplitude (shown by dashed curve)  of the soliton solution \eqref{eq:KdV5_soliton_solution} with the radius of bubbles in the unperturbed state ($R_{0}$) for water containing carbon dioxide bubbles with $C_{0}=1$ and $\phi_{0}$=0.001. }
  \label{fig:2.1}
\end{figure}

From \eqref{eq:KdV5_solitary_wave_speed} we see that the amplitude of the solitary wave depends on the wave speed and parameters $\alpha$ and $\varepsilon$. The width of the solitary wave depends on the wave speed and parameters $\beta$ and $\varepsilon$. Using values of $\alpha$, $\beta$ and $\varepsilon$ we can find variations of $v_{\mbox{\tiny{m}}}$ and $\Delta$ with radius of bubbles in the unperturbed state that are presented in Fig.\ref{fig:2.1}. From Fig.\ref{fig:2.1} we see that the amplitude $v_{\mbox{\tiny{m}}}$ slowly decreases when the radius of bubbles in the unperturbed state increases. The width of the solitary waves $\Delta$ rapidly increases when the equilibrium bubbles radius increases.

The generalized Korteweg--de Vries equation \eqref{eq:extended_KdV_equation} and its particular cases have been studied by the present time. Let us consider the general case of \eqref{eq:extended_BKdV_equation_3}. Equation \eqref{eq:extended_BKdV_equation_3} is new and it is interesting to construct exact solutions for this equation.

It is convenient to use in \eqref{eq:extended_BKdV_equation_3} the values of  $\lambda_{1},\lambda_{2},\lambda_{3}$ the same as for the purely dispersive case (equation \eqref{eq:extended_KdV_equation}), because we see that equations \eqref{eq:extended_KdV_equation}, \eqref{eq:extended_BKdV_equation_3} have the same leading order terms. Thus, using \eqref{eq:KdV5_parameters} we guarantee a simple form for the Laurent series of the solution.

Using \eqref{eq:KdV5_parameters} in \eqref{eq:extended_BKdV_equation_3}
we obtain
\begin{equation}
\begin{gathered}
v_{t}+ \alpha v v_{x}+ \beta v_{xxx}-\mu v_{xx}+ \varepsilon\Big[2\beta^{2} v_{xxxxx}+\frac{5\alpha^{2}}{3} v^{2}v_{x}+\frac{10}{3}\beta\alpha vv_{xxx}+\\+\frac{20}{3}\beta\alpha v_{x}v_{xx}-\nu_{1}vv_{xx}-\nu_{2}v_{x}^{2}+ \mu^{2} v_{xxx}+(\gamma-3\beta\mu) v_{xxxx}\Big]=0,
\label{eq:extended_BKdV_equation_4}
 \end{gathered}
\end{equation}
where
\begin{equation}
\begin{gathered}
\nu_{1}=\frac{(17\mu\beta\alpha+6\mu\beta_{1}+9\nu\beta)}{9\beta},\quad
\nu_{2}=\frac{(35\mu\alpha\beta+9\nu\beta-3\mu\beta_{1})}{9\beta}.
 \end{gathered}
\end{equation}

\begin{figure}
\center
 \includegraphics[width=0.55\textwidth]{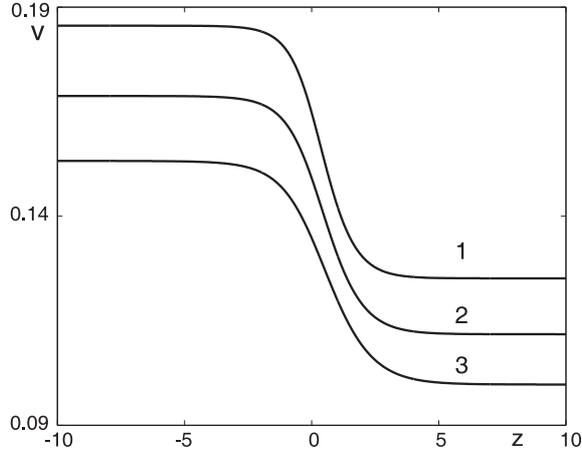}
\caption{ Variation of exact solution \eqref{eq:extBKdV_kink_solution} with $z$ for different values of radius of bubbles in the unperturbed state $R_{0}=0,125$mm (Curve 1), $R_{0}=0.133$mm (Curve 2) and $R_{0}=0.142$mm (Curve 3) for water containing carbon dioxide bubbles with $\phi_{0}=0.001$.}
  \label{fig:2}
\end{figure}

Equation \eqref{eq:extended_BKdV_equation_4} admits elliptic, solitary wave and rational solutions. To construct these solutions we can use the simplest equation method \cite{Kudryashov2005}.
For example equation \eqref{eq:extended_BKdV_equation_4} has a solution in the form
\begin{equation}
\begin{gathered}
v(x,t)=A_{0}+A_{1}\sqrt{B}\tanh\{\sqrt{B}(z-z_{0})\}
-\frac{36\beta}{\alpha}\,B\tanh^{2}\{\sqrt{B}(z-z_{0})\},\\ z=x-C_{0}\,t,
\label{eq:extBKdV_kink_solution}
 \end{gathered}
\end{equation}
where
\begin{equation}
\begin{gathered}
A_{1}=-\frac{2(12\mu\beta_{1}+76\mu\alpha\beta+45\nu\beta+15\gamma\alpha)}{35\beta\alpha^{2}},\vspace{0.1cm}\\
\label{eq:extBKdV_kink_solution_p1}
 \end{gathered}
\end{equation}
$z_{0}$ is an arbitrary constant. We do not present expressions for $A_{0},C_{0}$ and $B_{0}$ due to their cumbersome form. There is a correlation on parameters $\alpha,\beta,\beta_{1},\mu,\nu_{1},\nu_{2}$ for the existence of solution \eqref{eq:extBKdV_kink_solution}. We do not present this correlation either for the same reason, however we note that this correlation holds for physically relevant values of parameters $\alpha,\beta,\beta_{1},\mu,\nu_{1},\nu_{2}$.

The plots of solution \eqref{eq:extBKdV_kink_solution} are presented in Fig.\ref{fig:2} for water with carbon dioxide bubbles of various radiuses and at $z_{0}=0$. From Fig.\ref{fig:2} we see that solution \eqref{eq:extBKdV_kink_solution} is a weak shock wave. The amplitude of this wave decreases with the increasing of bubbles radius in the unperturbed state. The width of shock wave \eqref{eq:extBKdV_kink_solution} increases under the same condition. Let us note that nonlinear waves similar to presented in Fig.\ref{fig:2} were observed experimentally (e.g. see \cite{Nakoryakov1993,Nigmatulin1990,Wijngaarden1974}). The other solutions of \eqref{eq:extended_BKdV_equation_4} have to be constructed elsewhere.

\section{Conclusion}

We have studied long weakly nonlinear waves in the liquid containing gas bubbles. We have obtained equations for nonlinear waves taking into account high order terms in the asymptotic expansion (equations \eqref{eq:ext_Burgers_1}, \eqref{eq:extended_BKdV_equation_1}). These equations govern  nonlinear waves at different length and time scales and are the generalizations of the Burgers equation and the Burgers--Korteweg--de Vries equation correspondingly.

We have found normal form equation \eqref{eq:ext_Burgers_3} for \eqref{eq:ext_Burgers_1} using near--identity transformations. We have shown that this normal form equation is integrable under a certain condition on physical parameters (formula \eqref{eq:STO_parameters1}). It is worth noting that condition \eqref{eq:STO_parameters1} is realized for the liquid with gas bubbles. Thus, we have found that equation \eqref{eq: main_evolution_equation} is asymptotically integrable at $m=1$ with accuracy up to  $\varepsilon$ under condition \eqref{eq:STO_parameters1} for waves in the gas--liquid mixture. In the general case of \eqref{eq:ext_Burgers_3} we have obtained and analyzed a solitary wave solution.

The generalizations of the Burgers--Korteweg--de Vries equation were analyzed by near--identity transformations as well. We have found normal form equation \eqref{eq:extended_BKdV_equation_3}. We have shown that in the purely dispersive case \eqref{eq:extended_BKdV_equation_3}  is the generalized Korteweg--de Vries equation. Using arbitrary parameters $\lambda_{1},\lambda_{2},\lambda_{3}$ from near--identity transformations \eqref{eq:Kodama_transformations2} we can transform the generalized Korteweg--de Vries equation to one of the integrable fifth order evolution equations. Thus, dynamics of nonlinear waves in the liquid with gas bubbles is asymptotically integrable in the purely dispersive case with accuracy up to $\varepsilon$. We have found exact special solutions of \eqref{eq:extended_BKdV_equation_3} in the general case.

Let us note that system of equation \eqref{eq:main_non_dym_system} used in the current work for the investigation of nonlinear waves in the liquid containing gas bubbles can be used for the investigation of nonlinear waves in other media. For example, one can use this system of equations for waves in visco--elastic tubes \cite{Kudryashov2006,Demiray2009}, nonlinear waves in plasma \cite{Washimi1966} and other applications \cite{Leblond2008}. Thus, the results of our work can be applied not only to the study of waves in the liquid with gas bubbles but also to the investigation of nonlinear waves in all media described by system of equations (5).

\section{Acknowledgments}

This research was partially supported by RFBR grant 14-01-00493-a, by grant for Scientific Schools 2296.2014.1. and by grant for the state support of young Russian scientists 3694.2014.1.

\appendix

\section{The nondimensional parameters of system of equations \eqref{eq:main_non_dym_system}}
\label{sec:A}
The nondimensional parameters of system of equations \eqref{eq:main_non_dym_system} have the form

\begin{subequations}
\begin{equation}
\tilde{\alpha} = \frac{3\chi\rho_{0}}{R_{0}}-\frac{2\sigma\chi\rho_{0}}{R_{0}^{2}P_{0}},
\end{equation}
\begin{align}
& \tilde{\beta}=\frac{\rho_{l} \chi R_{0} c_{0}^{2} \rho_{0}}{P_{0}\,L^{2}}+\frac{\tilde{\lambda}_{4}\rho_{0}\chi(\gamma_{g}-1)^{2}}{525R_{0}\gamma_{g}^{2}D^{2}}, \end{align}
\begin{align}
& \tilde{\alpha}_{1}=\frac{(6\chi^{2}-3\,\chi_{1}\,R_{0})\rho_{0}^{2}}{R_{0}^{2}}
    -\frac{2\sigma(\chi^{2}-\chi_{1}R_{0})\rho_{0}^{2}}{P_{0}R_{0}^{3}}
\end{align}
\begin{align} \tilde{\beta}_{1}=\frac{2(\gamma_{g}-1)^{2}(\chi_{1}R_{0}-3\chi^{2})\rho_{0}^{2}}{525\gamma_{g}^{2}R_{0}^{2}D^{2}}
+ \frac{\rho_{l}(2 \chi_{1} R_{0}+\chi^{2}) c_{0}^{2} \rho_{0}^{2}}{P_{0}\,L^{2}},
\end{align}
\begin{align} &\tilde{\beta}_{2}=\frac{(\gamma_{g}-1)^{2}[(\tilde{\lambda}_{3}-2\tilde{\lambda}_{4})\chi^{2}+
    2\chi_{1}\tilde{\lambda}_{4}R_{0}]\rho_{0}^{2}}{525\gamma_{g}^{2}R_{0}^{2}D^{2}}    +\frac{\rho_{l}(4\chi_{1}R_{0}\chi^{2}+3\chi^{2}) \rho_{0}^{2} c_{0}^{2}}{2P_{0}\,L^{2}},
\label{eq: non-dim_parameters1}
\end{align}
\begin{align}
& \tilde{\mu}=\frac{(\gamma_{g}-1)\rho_{0}\chi}{5\gamma_{g} R_{0}D}+\frac{4\nu_{l}\chi\,\rho_{l}\rho_{0}c_{0}}{3R_{0}P_{0}L},
\end{align}
\begin{align}
&\tilde{\mu}_{1}=\frac{(\gamma_{g}-1)(5\chi^{2}-2\chi_{1}R_{0})\rho_{0}^{2}}{5\gamma_{g} R_{0}^{2}D}+\frac{(4\nu_{l}\chi^{2}-8\nu_{l}\chi_{1}R_{0})\rho_{l}\rho_{0}^{2}c_{0}}
    {3R_{0}^{2}P_{0}L},
\end{align}
\begin{align}
& \tilde{\gamma}=\frac{\chi R_{0}^{2}\rho_{l}\rho_{0}c_{0}^{3}}{c_{l}P_{0}L^{3}},
\end{align}
\begin{align}
& \tilde{\gamma}_{1}=\frac{2R_{0}\rho_{l}\rho_{0}^{2}c_{0}^{3}}{P_{0}c_{l}L^{3}}\Bigg[\chi_{1}R_{0}+\chi^{2}\Bigg],
\end{align}
\begin{align}
&D=\frac{\chi_{g} }{\tilde{\omega} R_{0}^{2}},\quad \tilde{\omega}=\frac{c_{0}}{L}.
     \label{eq: non-dim_parameters2}
\end{align}
\label{eq: non-dim_parameters}
\end{subequations}

We use in \eqref{eq: non-dim_parameters} the following notations:
\begin{equation}
\tilde{\lambda}_{1}=\frac{(\gamma_{g}-1)R_{0}^{3}}{5\gamma_{g}\chi_{g}},\quad
\tilde{\lambda}_{2}=\frac{(\gamma_{g}-1)^{2}R_{0}^{6}}{525\gamma_{g}^{2}\chi_{g}^{2}}, \quad
\tilde{\lambda}_{3}=(2+15 K_{0}^{'}),\quad \tilde{\lambda}_{4}=\frac{12\gamma_{g}-7}{3(\gamma_{g}-1)}.
\end{equation}

Nondimensional parameter $\alpha_{1}$ correspond to the nonlinearity of waves, parameters $\tilde{\mu},\tilde{\mu}_{1},\tilde{\gamma},\tilde{\gamma}_{1}$ correspond to the dissipation of the nonlinear waves, parameters $\tilde{\beta},\tilde{\beta}_{1},\tilde{\beta}_{2}$ correspond to the dispersion of the nonlinear waves.

\section{The parameter of solution \eqref{eq:ext_Burgers_3_solution}}
\label{sec:B}
The parameter $B$ in solution \eqref{eq:ext_Burgers_3_solution} has the form
\begin{equation}
\begin{gathered}
B= \left( 4\,\mu+\alpha \right)\bigg[ \alpha\left( 2\mu+\alpha \right)\left( 4
\mu+\alpha \right)  \left( 4\,\mu^{3}+\mu\alpha^{2}+4\mu
\beta-6\,\mu\nu+6\beta\alpha+\nu\alpha \right)-\hfill \\-16\varepsilon \big( 3\beta\alpha+3\mu^{2}\alpha+\mu\alpha^{2}+\nu\alpha+4\mu\beta+4\mu^{3} \big)^{2}C_{0}  \bigg] \times \hfill \\
\times \bigg[16\varepsilon^{2}\left(3\,\beta\,\alpha+3\,\mu^{2}\alpha+\mu\alpha^{2}+\nu\alpha+4\,\mu\,\beta+4\mu^{3}\right) ^{2}\left( 2\,\beta+2\,{\mu}^{2}+\mu\,\alpha+\nu \right) \alpha\bigg]^{-1}.
\end{gathered}
\end{equation}



\end{document}